\documentclass[aps,prc,twocolumn,superscriptaddress]{revtex4-1}

\usepackage[dvipsnames]{xcolor}
\usepackage{tikz}
\usetikzlibrary{shapes.misc}

\usepackage{amsmath,amssymb,graphicx}
\usepackage{soul,xcolor}
\usepackage{url}
\usepackage[colorlinks,urlcolor=blue]{hyperref}

\begin{document}

\title{Flavor dependence of Energy-energy correlators}

\author{Liliana Apolinário}
\affiliation{LIP, Av. Prof. Gama Pinto, 2, P-1649-003 Lisboa, Portugal}
\affiliation{Instituto Superior Técnico (IST), Universidade de Lisboa,  Av. Rovisco Pais 1, P-1049-001 Lisboa, Portugal}

\author{Raghav Kunnawalkam Elayavalli}
\affiliation{Department of Physics and Astronomy, Vanderbilt University, Nashville, TN}

\author{Nuno Olavo Madureira}
\affiliation{LIP, Av. Prof. Gama Pinto, 2, P-1649-003 Lisboa, Portugal}
\affiliation{Instituto Superior Técnico (IST), Universidade de Lisboa,  Av. Rovisco Pais 1, P-1049-001 Lisboa, Portugal}

\author{Jun-Xing Sheng}
\affiliation{Department of Physics and Astronomy, Vanderbilt University, Nashville, TN}

\author{Xin-Nian Wang}
\affiliation{Key Laboratory of Quark and Lepton Physics (MOE) \& Institute of Particle Physics, Central China Normal University, Wuhan 430079, China}

\author{Zhong Yang}
\affiliation{Department of Physics and Astronomy, Vanderbilt University, Nashville, TN}
\affiliation{Key Laboratory of Quark and Lepton Physics (MOE) \& Institute of Particle Physics, Central China Normal University, Wuhan 430079, China}

\begin{abstract}
Energy-energy correlators (EECs) within high energy jets serve as a key experimentally accessible quantity to probe the scale and structure of the quark-gluon plasma (QGP) in relativistic heavy-ion collisions. The CMS Collaboration's first measurement of the modification to the EEC within single inclusive jets in Pb+Pb collisions relative to p+p collisions reveals a significant enhancement at small angles, which may arise from jet transverse momentum $p_T$ selection biases due to jet energy loss. We investigate the dependence of jet EECs on the flavor of the initiating parton. The EEC distribution of a gluon jet is broader and the peak of transition from the perturbative to the non-perturbative regime occurs at a larger angle than that of a quark jet.  Such flavor dependence leads to the different EECs for $\gamma$-jets and single inclusive jets due to their different flavor composition. It is also responsible for a colliding energy dependence of EECs of single inclusive jets at fixed jet energy. We also investigate the impact of flavor composition variation on the $p_T$ dependence of the jet EEC. We further propose that a change in the gluon jet fraction in A+A collisions compared to p+p can also contribute to a non-negligible enhancement of the medium modified EEC at small angles. Using the \textsc{jewel} model, we predict the reduction of the gluon jet fraction in A+A collisions and estimate its impact on the EEC. 
\end{abstract}

\maketitle

\noindent{\it \color{blue} Introduction --}
Jet is a cluster of energetic particles moving in the same direction~\cite{Sterman:1977wj}. It is formed by the fragmentation of a high-virtuality parton (quark or gluon) produced in the initial hard scattering. When a jet traverses the Quark-Gluon Plasma (QGP), it undergoes interactions with the medium~\cite{Bjorken:1982tu, Thoma:1990fm, Braaten:1991we, Gyulassy:1993hr, Baier:1996kr, Zakharov:1996fv, Gyulassy:1999zd, Wiedemann:2000za, Wang:2001ifa, Arnold:2002ja, Djordjevic:2006tw, Qin:2007rn, Apolinario:2022vzg}, carrying important information of the medium's properties embedded in the momentum exchange with the medium constituents.
Therefore, jets act as powerful hard probes of the QGP properties. Jet physics has experienced significant development over the past three decades. Jet quenching caused by the energy loss of jets in high-energy heavy-ion collisions~\cite{Gyulassy:2004zy, Wang:2004dn, Apolinario:2024equ} is a key experimental observable to prove the existence of QGP at both RHIC and LHC energies~\cite{PHENIX:2001hpc, STAR:2002ggv, ATLAS:2010isq, ALICE:2010yje, CMS:2011iwn}. Nowadays, there is growing interest in the study of jet substructure to understand the space-time structure of the parton shower, encompassing a range of more precise measurements, such as jet shape~\cite{Chien:2015hda, CMS:2018jco, Luo:2018pto, Chang:2019sae, Yang:2022nei}, jet fragmentation function~\cite{Casalderrey-Solana:2016jvj, CMS:2018mqn, ATLAS:2018bvp, Chen:2020tbl}, groomed jets~\cite{Chien:2016led, ALargeIonColliderExperiment:2021mqf} and jet reclustering~\cite{Apolinario:2020uvt, Apolinario:2024hsm}.

Among the various jet substructure observables, the Energy-Energy Correlator (EEC)~\cite{Basham:1977iq, Basham:1978bw, Basham:1978zq} stands out as a particularly powerful one. It is based on the correlations of asymptotic energy flux deposited in detectors~\cite{Chen:2020vvp, Komiske:2022enw}, offering significant potential for probing the internal structure of jets across a wide range of scenarios~\cite{Liu:2022wop, Liu:2023aqb, Cao:2023oef, Chen:2019bpb, Chen:2020adz, Chen:2022swd, Holguin:2022epo, Lee:2022uwt, Devereaux:2023vjz, Jaarsma:2023ell, Lee:2023npz, Kang:2023big, Chen:2024nfl, Holguin:2023bjf, Bossi:2024qho, Singh:2024vwb, Alipour-fard:2025dvp, Craft:2022kdo, Xing:2024yrb, Andres:2023ymw}.
It has been found that the EEC can not only be used to probe the onset of color coherence in in-medium splittings~\cite{Andres:2022ovj, Andres:2023xwr}, but also shed light on the dead-cone effect and the flavor hierarchy of heavy flavor jets~\cite{Craft:2022kdo, Andres:2023ymw, Xing:2024yrb, alicehfjet}. Furthermore, the modification of jet constituents by QGP medium will ultimately manifest in the angular distribution of the EEC within the jet. Therefore,  both theoretical and experimental efforts are focused on the differences in EEC within jets between Pb+Pb(A+A) and p+p collisions, exploring the connection between medium modification and the scale and structure of the QGP medium~\cite{Andres:2022ovj, Andres:2023xwr, Andres:2023ymw, Yang:2023dwc, Barata:2023bhh, Andres:2024ksi, Bossi:2024qho, Singh:2024vwb, Andres:2024hdd, Andres:2024pyz, Xing:2024yrb, Barata:2024ieg, Chen:2024cgx, Barata:2024nqo, Andres:2024xvk,  CMS:2024ovv}. In addition, EEC is also expected to be sensitive to cold nuclear matter effects in small systems~\cite{Andres:2024xvk, Fu:2024pic, Barata:2024wsu}. 

Recently, the CMS collaboration released the first measurement of the EEC within single inclusive jets in Pb+Pb collisions at $\sqrt{s_{\rm NN}}=5.02$ TeV~\cite{jussi}. Their data indicates that the EEC distribution in Pb+Pb collisions shows a significant enhancement at both small and large angle relative to p+p collisions. The enhancement at large angles has been predicted by multiple theoretical studies, which generally attribute it to contributions from jet-induced medium response, medium-induced gluon radiation and transverse momentum broadening~\cite{Andres:2024ksi, Andres:2022ovj, Andres:2023xwr, Yang:2023dwc,Xing:2024yrb}. However, the enhancement at small angles remains poorly understood. One potential explanation is that it arises from the jet $p_T$ selection bias ~\cite{Andres:2024hdd, Xing:2024yrb, Chen:2024cgx, STAR:2025jut} due to jet energy loss. In experimental measurements, using the same reconstructed $p_T$ for jets in A+A and p+p collisions implies that the initial jet $p_T$ in A+A collisions is higher than that in the corresponding p+p collisions. Since the transition region of the EEC distribution depends on $\lambda_{\rm QCD}/p_T^{\rm jet}$~\cite{Komiske:2022enw, jussi, CMS:2024ovv, ALICE:2024dfl, Stareec}, the EEC distribution within the initial jet in A+A events will shift toward smaller angles relative to p+p collisions. Under the assumption that the EEC of the final jet in A+A collisions is determined by the initial jet energy scale, this shift could result in the observed enhancement of self-normalized EEC distributions in A+A collisions at small angles when compared to p+p collisions. This shift has also been addressed recently with a quantitative method to reduce this bias in data~\cite{Apolinario:2024apr, Andres:2024hdd}.


In this \emph{Letter}, we explore the dependence of the jet EEC on the flavor of the initiating parton, or simply referred to as the flavor dependence, and the relevant consequences in the nuclear modification of EEC in heavy-ion collisions. We calculate the EEC within quark and gluon initiated jets separately in p+p collisions and show that the variation in the gluon jet fraction between A+A and p+p collisions is a non-negligible factor contributing to the enhancement of the EEC at small angles. Using \textsc{pythia 8.3} to describe vacuum physics and the \textsc{jewel 2.3} model to simulate medium-induced effects, we estimate the changes in the gluon jet fraction between p+p and A+A and predict their impact on the EEC distribution in realistic heavy-ion collisions.

\noindent{\it \color{blue} EECs of quark and gluon jets--}
Jets initiated by high-energy gluons and quarks are expected to exhibit distinct EEC distributions due to the differences in their color charges and fragmentation patterns. 
\begin{figure}[h!]
\centering
	\includegraphics[width=0.5\textwidth]{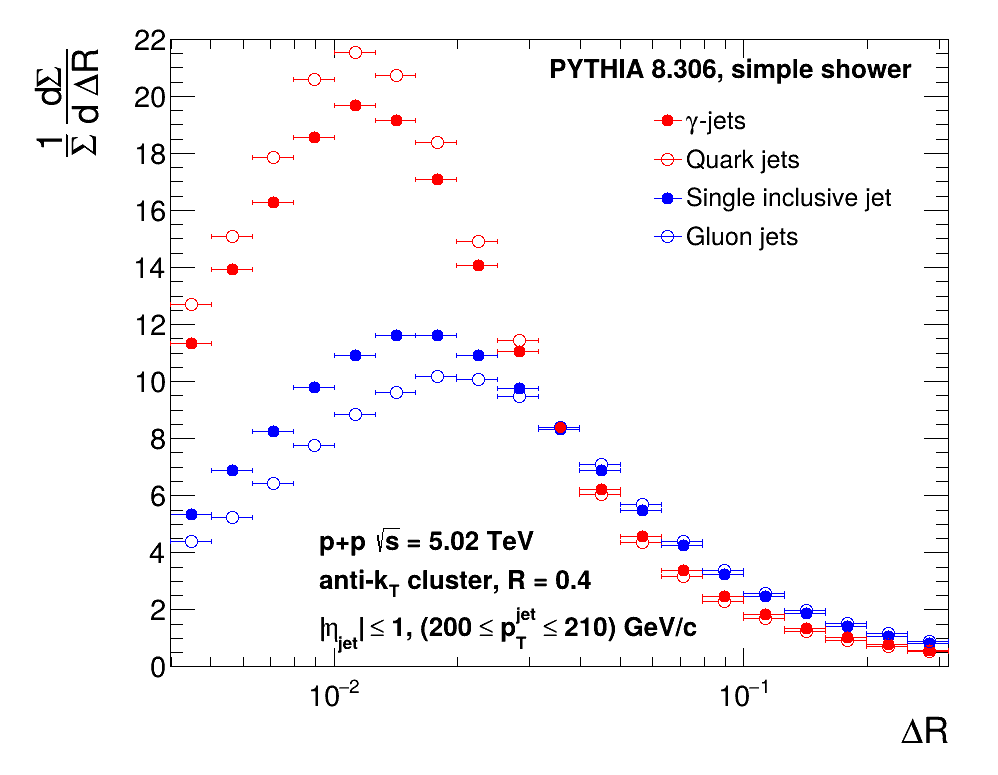}
	\caption{Angular distribution of the EEC for quark jets (red open circle), gluon jets (blue open circle), $\gamma$-jets (red filled circle) and single inclusive jets (blue filled circle) in p+p collisions generated by \textsc{pythia8}.}
	\label{qgjet}
\end{figure}
Fig.\ref{qgjet} shows the EEC distribution within gluon jets (blue open circle) and quark jets (red open circle) in p+p collisions. In this \emph{Letter}, unless otherwise specified, all EECs are self normalized to 1. Jets are reconstructed using \textsc{fastjet}~\cite{Cacciari:2011ma} with the anti-$k_T$ algorithm~\cite{Cacciari:2008gp} with the jet radius $R=0.4$. Vacuum events are generated using the \textsc{pythia8}~\cite{Sjostrand:2007gs, Sjostrand:2014zea, Bierlich:2022pfr} Monte Carlo event generator with the default tune. 
We require both jet populations to have the same reconstructed $p_T$ range and restrict them to a narrow interval (200-210) GeV/c, to reduce any $p_T$ spectra shape effects.
We find that gluon jets tend to have a significantly broader distribution of EEC compared to those particle pairs within quark jets. This effect can be attributed to several key features of parton evolution in QCD. First, the higher number of gluon color charges allows for a greater variety of splitting processes compared to quarks. Second, the color factor for gluons, $C_A$ = 3, is significantly larger than the quark color factor, $C_F$ = $\frac{4}{3}$, leading to a higher probability of emissions. Meanwhile, in \textsc{pythia} the hadronization scale is set by the parton's virtuality $Q^2$ which is related to the transverse momentum of the splitting $k_T^2=z(1-z)Q^2$. Due to the symmetrical splitting functions of gluons, the average $k_T$ or the relative angle of final partons in a gluon jet is slightly larger than a quark jet.
Consequently, gluons generate more particles within the jet, resulting in a broader EEC distribution with its peak shifted toward larger angles. Such a phenomenon can also be observed when comparing the EEC within single inclusive jets (or di-jets) to that of $\gamma$-jets. In Fig.\ref{qgjet}, we also plot the EEC within $\gamma$-jets (red filled circle) and single inclusive jetS (blue filled circle) in p+p collisions. 
As compared to results of quark jets and gluon jets, we find that the EEC of $\gamma$-jets is closer to that of quark jets, while the result of single inclusive jets is closer to that of gluon jets. This indicates that at LHC energies, within the current jet $p_T$ range, $\gamma$-jets are dominated by quark jets, whereas single inclusive jets are dominated by gluon jets. 

Since the flavor composition of single inclusive jets varies with the initial parton momentum fraction $x=2p_T/\sqrt{s}$~\cite{Liu:2024lxy}, the flavor dependence of jet EEC should lead to a colliding energy dependence of single inclusive jet EEC with same jet $p_T$.
\begin{figure}[h!]
    \centering
    \includegraphics[width=0.5
    \textwidth]{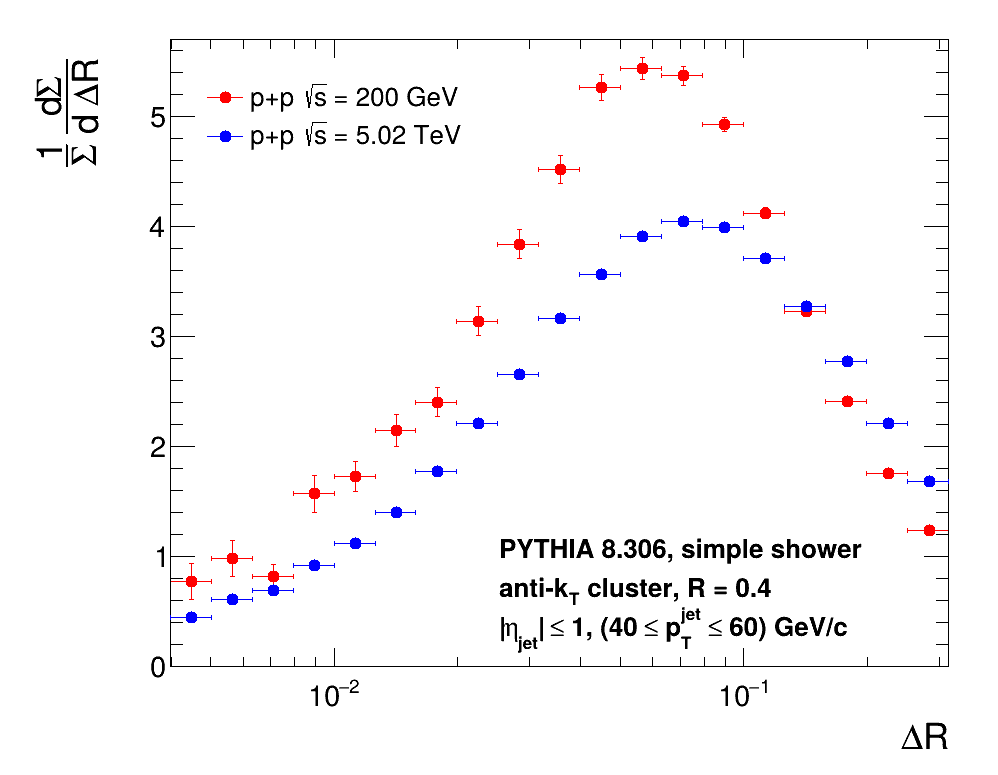}
	\caption{EEC within single inclusive jets with the same jet $p_T$ in p+p collisions at RHIC (red) and LHC (blue), respectively.}
	\label{rhiclhc}
\end{figure}
In Fig.~\ref{rhiclhc}, we present the EEC of single inclusive jets with $p_T^{jet} \in$ (40-60) GeV/c in p+p collisions at RHIC (red) and LHC (blue). 
We choose this specific jet $p_T$ range at RHIC since it is accessible in the recent data-sets from sPHENIX and ongoing measurements by STAR. It is also accessible by ALICE at LHC, providing a study of the $x$ dependence of jet EECs. 
We observe that the EEC at the RHIC energy shifts to smaller angles compared to that at LHC energy, since for the same jet $p_T$, single inclusive jets at the RHIC energy have a larger initial momentum fraction $x$ and therefore a larger fraction of quark jets. As a result, the EEC distribution within single inclusive jets at RHIC resembles more the quark jet EEC than the one for single inclusive jets at LHC.

\begin{figure}[h!]
\centering
	\includegraphics[width=0.48
    \textwidth]{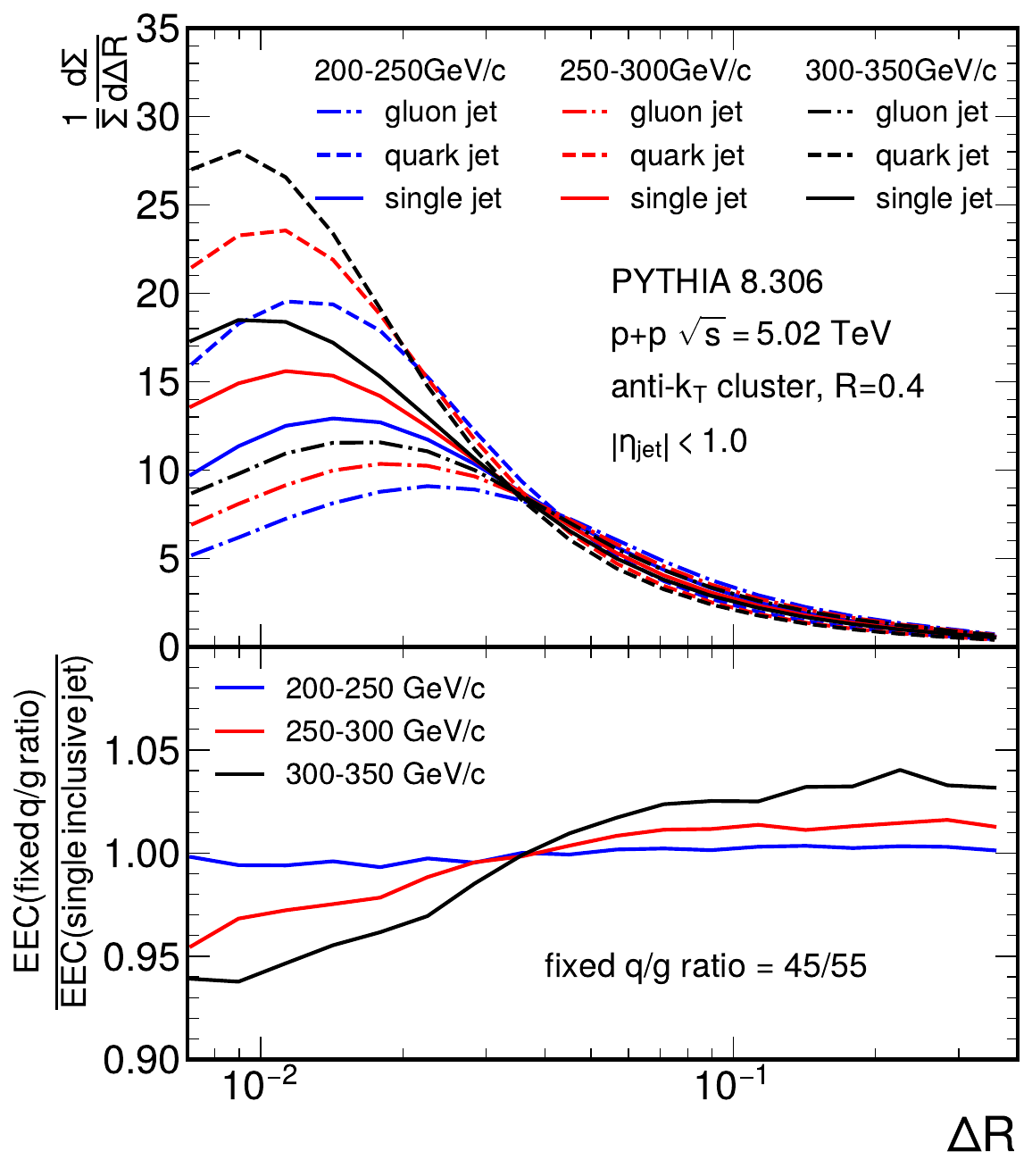}
	\caption{EEC distribution (upper panel) of quark jets (dashed), single inclusive jets (solid) and gluon jets (dot-dashed) for 3 jet $p_T$ ranges: (200-250) GeV/c (blue), (250-300) GeV/c (red) and (300-350) GeV/c (black) in p+p collisions and the ratio (lower panel) of the single jet EEC with fixed gluon jet fraction to the default single jet EEC distribution for the same jet $p_T$ ranges (same color scheme).}
	\label{flavor_ptbias}
\end{figure}
When accounting for the $p_T$ dependence of jet EEC, one must also consider the impact of jet flavor variation, as the flavor composition changes with jet $p_T$.
To illustrate such $p_T$ dependence of jet EECs due to changes of flavor composition, we present the EEC distributions for inclusive jets, quark jets, and gluon jets in three different jet $p_T$ intervals in the upper panel of Fig.~\ref{flavor_ptbias}. We observe clearly different $p_T$ dependence of quark and gluon jet EECs. The $p_T$ dependence of single inclusive jet EECs is a combination of the $p_T$ dependence of quark and gluon jet EECs and the flavor composition. For a given jet $p_T$ range, one can use the quark (dashed lines) and gluon (dot-dashed lines) jet EEC distributions to construct the single inclusive jet EEC (solid lines) and extract the flavor composition. For example, the $q/g$ fraction is approximately 45$\%$/55$\%$ at $p_T^{jet} \in$ (200-250) GeV/c using this method.

To quantify the effect of jet flavor variation on the $p_T$ dependence of single inclusive jet EEC, we construct the single inclusive jet EEC for each $p_T$ range with a fixed quark and gluon jet ratio of 45$\%$/55$\%$ and compare to the single inclusive jet EECs with the flavor composition given by \textsc{pythia}.
In the lower panel of Fig.~\ref{flavor_ptbias}, we plot the ratio of the constructed EEC distributions with fixed quark/gluon ratio to the single inclusive jet EECs from \textsc{pythia} at the same jet $p_T$. We observe that this ratio for $p_T^{jet} \in$ (200-250) GeV/c (blue) is approximately equal to 1, indicating that the single jet EEC can be well described as a combination of the quark and gluon jet EECs. As the jet $p_T$ increases, this ratio falls below 1 at small angles as the fraction of gluon jets gradually decreases with increasing jet $p_T$, causing the single jet EEC to shift closer to the quark jet EEC. This trend becomes more pronounced at higher jet $p_T$.

\begin{figure}[h!]
\centering
	\includegraphics[width=0.5\textwidth]{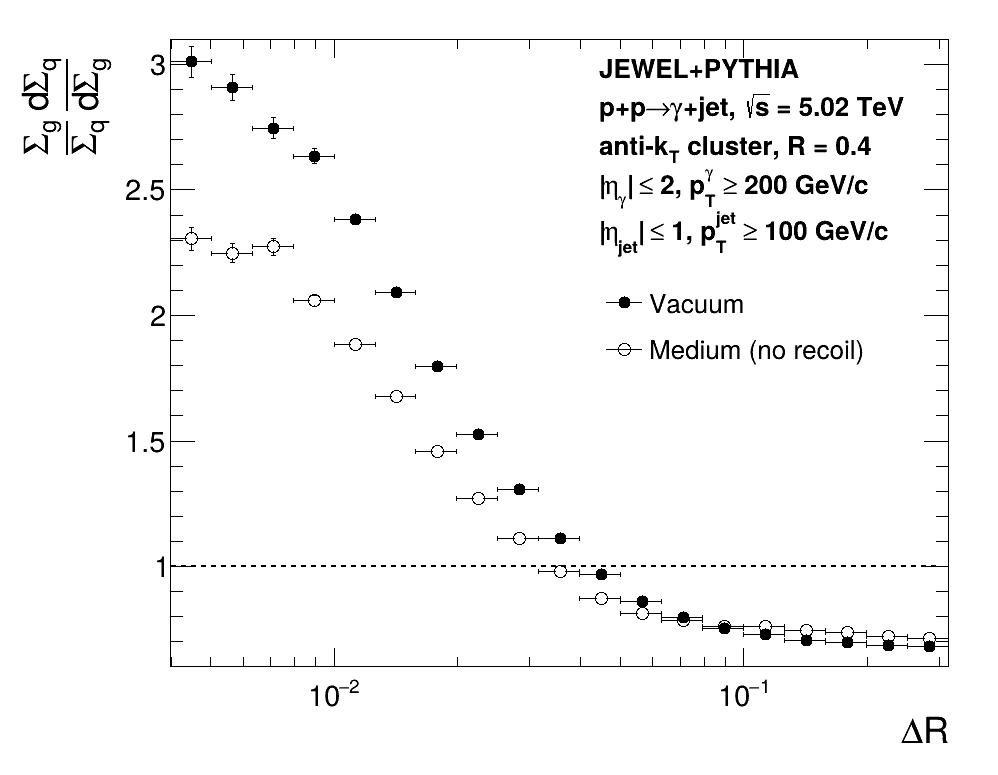}
	\caption{Ratio of the EEC distribution inside quark jets to that inside gluon jets from p+p collisions (full circle) and after quenching within the \textsc{jewel} model (open circle).}
	\label{EEC_vac_med}
\end{figure}

\noindent{\it \color{blue} Medium modification of small-angle jet EEC --}
For A+A collisions, the difference in Casimir factors causes gluon jets to lose more energy compared to quark jets~\cite{Apolinario:2020nyw, Qin:2015srf}. As a result, the EEC within gluon jets exhibits a stronger $p_T$ selection bias when compared to quark jets. This bias effectively reduces the difference in EEC between quark jets and gluon jets in A+A collisions compared to p+p collisions. 
To simulate effects of medium modification, we use \textsc{jewel} 2.3 model~\cite{Zapp:2012ak, KunnawalkamElayavalli:2016ttl, Milhano:2022kzx}, with an initial mean temperature of $T_i = 550~\rm{MeV}$ and a $[0-10]\%$ centrality class following~\cite{Milhano:2022kzx}. The remaining parameters that model the boost-invariant longitudinal expansion of the ideal quark-gluon gas implemented in \textsc{jewel} are left to its default values. Finally, we restrict ourselves to energy loss effects and disregard the medium response that would populate the large $\Delta R$ region of the EEC. With this setup, we simulate independently $\gamma + q-$ and $\gamma + g-$initiated jets and present their ratio of EEC for both vacuum (filled circle) and medium (open circle) configurations in Fig.\ref{EEC_vac_med}. Our result shows that the medium effect indeed diminishes the difference in EEC between quark jets and gluon jets at small angles, reducing the ratio from 3.0 to approximately 2.3.

Furthermore, the greater energy loss experienced by gluon jets compared to quark jets also reduces the fraction of gluon jets in the final state in A+A collisions relative to p+p collisions. This shift makes the contribution of quark jets to the EEC more pronounced in A+A collisions. 
\begin{figure}[h!]
\centering
	\includegraphics[width=0.5\textwidth]{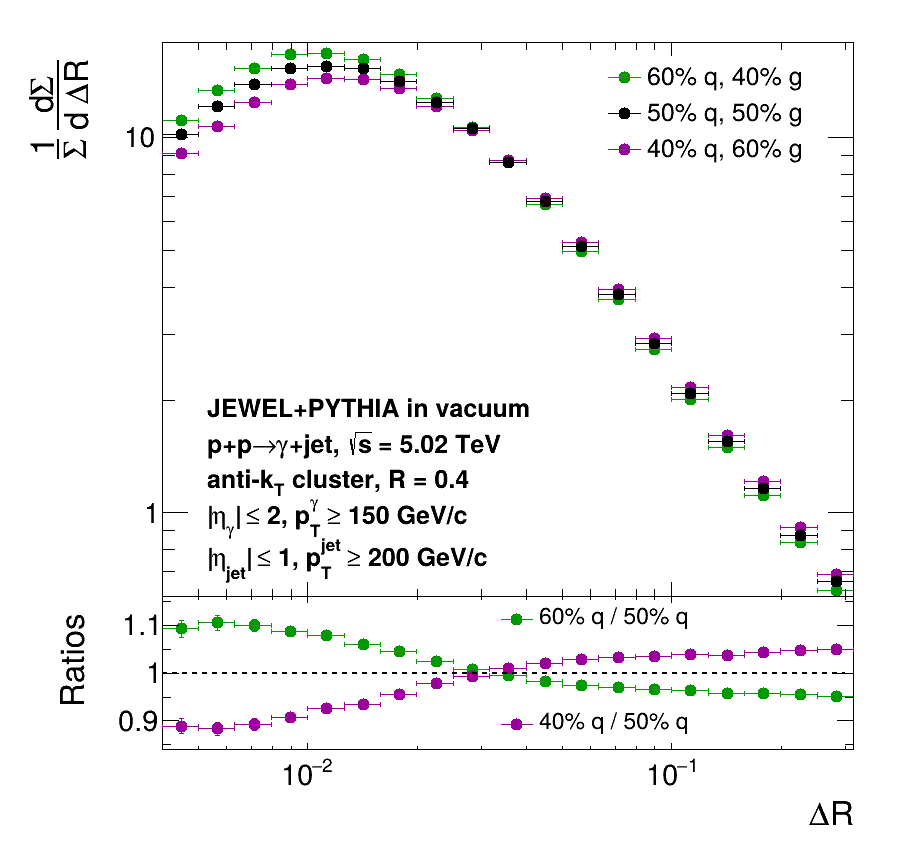}
	\caption{EEC distributions in p+p collisions (upper panel) with a 40\% (green), 50\% (black) and 60\% (purple) gluon jet fraction and ratio of the EEC for different gluon jet fractions compared to the 50$\%$ gluon jet fraction (lower panel).}
	\label{frag_qg}
\end{figure}
To better quantify this effect, in the upper panel of Fig. \ref{frag_qg}, we present the EEC distributions corresponding to different fractions of quark jets in p+p collisions. We observe that as the fraction of gluon jets decreases, the overall distribution shifts toward smaller angles. Additionally, we use the EEC with a 50$\%$ gluon jet fraction as the baseline and calculate the ratios of the EEC distributions for different gluon jet fractions relative to it in the lower panel of Fig. \ref{frag_qg}.
We find that a 10$\%$ change in the gluon jet fraction almost leads to a 10$\%$ variation in the EEC distribution at small angles. 
Based on \textsc{pythia} simulations of di-jet events with jet transverse momentum in the range of $p_T^{\rm{jet}} \in$ (120-140) GeV/c, the initial gluon jet fraction is approximately 66$\%$\footnote{This fraction depends on the PDF we used in the \textsc{pythia} model. Moreover, since we focus only on light quarks here, the contribution of heavy quarks is also excluded.}. However, after accounting for jet quenching effects within the \textsc{jewel} model, this fraction decreases slightly to 64.5$\%$ . 
As such, this medium-induced flavor selection can contribute approximately 6$\%$ of the enhancement of the EEC at small angles observed at LHC energies, making it a non-negligible effect when considering the overall QGP-induced modifications.

\noindent{\it \color{blue} Summary --} 
In this \emph{Letter}, we conduct a detailed investigation into the underlying mechanisms driving the enhancement of the measured single inclusive jet EEC at small angles in A+A collisions compared to p+p collisions. We primarily focus on the flavor dependence of the EEC and find that, in addition to the jet $p_T$ selection bias, the difference in the fraction of gluon jets between A+A and p+p collisions also plays a significant role. 
We highlight the key difference between di-jets and $\gamma-$jets in the EECs primarily due to the impact of jet-initiating parton flavor such as gluon jets in the former jet population and the latter having more quark jets.
We also discuss the potential comparison of similar jet momenta across RHIC and the LHC, noting the impact of the flavor variation on the small angle structure of the EEC. 
Using the \textsc{jewel} model, we estimate the reduction in gluon jet fraction in A+A collisions relative to p+p collisions and predict its specific non-negligible impact on the EEC distribution. 

We conclude that the EEC distribution within single inclusive jets is influenced not only by the  $p_T$  selection bias but also by the variation in the quark-to-gluon jet fraction in A+A compared to p+p collisions. The interplay of these effects results in a significant enhancement of the EEC ratio at small angles. In addition, this suggests that the EEC distribution within $\gamma$-jets may be different from that of single inclusive jets. The flavor dependence of the jet EEC also leads to $x$ dependence of jet EEC, for example, the colliding energy dependence for fixed jet $p_T$. We show some of the $p_T$ dependence of single inclusive jet EEC is also attributed to the variation of flavor composition. The consequences of the flavor dependence of jet EEC will influence a wide range of  studies  of jet structure and evolution in both p+p and A+A collisions. Our study here is just the beginning of further explorations.

\noindent{\it\color{blue} Acknowledgment --} 
This work is partially supported by the China Postdoctoral Science Foundation under Grant No. 2024M751059 and No. BX20240134(ZY), by NSFC under Grant No. 1193507 and by the Guangdong MPBAR with No.2020B0301030008. Computations in this study are performed at the NSC3/CCNU and NERSC under the
award NP-ERCAP0032607. RKE would like to thank the Vanderbilt ACCRE computing facility. RKE and JXS would also like to acknowledge funding by the U.S. Department of Energy, Office of Science, Office of Nuclear Physics under grant number DE-SC0024660. ZY and RKE are partially supported by the NSF XSCAPE program. LA and NOM were supported by FCT under contract 2021.03209.CEECIND and grant PRT/BD/154611/2022, respectively.

\bibliography{Ref}

\end{document}